\documentstyle[aps,prl,twocolumn,epsfig,graphics,amssymb]{revtex}

\draft
\begin{document}
\tighten
%%%%%%%%%%%%%%%%%%%%%%%%%%%%%%%
\onecolumn
\twocolumn[\hsize\textwidth\columnwidth\hsize\csname @twocolumnfalse\endcsname

\title{Modeling Supply Chains and Business Cycles as Unstable Transport Phenomena}
\author{Dirk Helbing$^{1}$} 

\address{$^1$ Institute for Economics and Traffic, 
Dresden University of Technology, Andreas-Schubert Str. 23, 01062 Dresden, Germany}

\maketitle
\begin{abstract}
Physical concepts developed to describe instabilities in traffic flows
can be generalized in a way that allows one to understand the well-known
instability of supply chains (the so-called ``bullwhip effect'').
That is, small variations in the consumption rate can cause
large variations in the production rate of companies generating
the requested product. Interestingly, the
resulting oscillations have characteristic frequencies
which are considerably lower than the variations in the
consumption rate. This suggests that instabilities of
supply chains may be the reason for the existence of business
cycles. At the same time, we establish some link to queuing theory and
between micro- and macroeconomics.\\
\mbox{ }
%The preconditions for these cycles are investigated in more detail.
\end{abstract}
]
%89.65.Gh Economics, business, and financial markets
%47.54.+r Pattern selection; pattern formation
%89.40.+k Transportation
%89.20.Bb Industrial and technological research and development
%%47.62.+q Flow control
%47.55.-t Nonhomogeneous flows
%47.70.-n Reactive, radiative, or nonequilibrium flows
%%89.20.-a Interdisciplinary applications of physics
%%89.75.-k Complex systems
%%89.75.Da Systems obeying scaling laws
%%89.75.Fb Structures and organization in complex systems
%%89.75.Hc Networks and genealogical trees
%%89.75.Kd Patterns

%\section{Introduction} 

Concepts from statistical physics and non-linear dynamics have been very successful in
discovering and explaining dynamical phenomena in traffic flows \cite{reviews}. Many of these phenomena
are based on mechanisms such as delayed adaptation to changing conditions and 
the competition for limited resources, which are relevant for other systems as well. This includes
pattern formation such as segregation in driven granular media \cite{segregation} and
lane formation in colloid physics \cite{colloid} or biological physics (pedestrians, ants) \cite{pedants}.
Another example are clogging phenomena at bottlenecks in freeway traffic \cite{phase}, panicking pedestrian crowds
\cite{panic}, or granular media \cite{clogging}. In the following study, we will focus on the phenomenon of
stop-and-go traffic \cite{stopgo} and its analogies. 
\par
Recently, economists and traffic scientists have wondered, whether 
traffic dynamics has also implications for the stability and management of supply chains 
\cite{Daganzo,Armbruster} or for the dynamics of business cycles \cite{Witt}. To explain business cycles, many
theoretical concepts have been suggested over the decades, 
such as the Schumpeter clock  \cite{business}.
These are usually based on
macroeconomic variables such as investment, income, consumption, public expenditure, or the
employment rate, and their interactions.  In contrast, Witt {\em et al.} \cite{Witt} have recently 
suggested to interpret business cycles as self-organization phenomenon due to a linear instability 
of production dynamics related to stop-and-go waves in traffic or driven many-particle systems. 
In order to illustrate their idea, they have transferred a continuous macroscopic
traffic model and re-interpreted the single terms and variables. 
%They have also carried out a linear stability
%analysis for another kind of fundamental diagram they consider to be realistic in their economic framework.
\par
The author believes that this is a very promising approach to understand business cycles, but
instead of simply transferring macroscopic traffic models, suggests 
to derive equations for business cycles from first principles, which means to derive the dynamics on
the macroscopic level from microscopic interactions. This would also make some contribution 
to the goal of understanding macroeconomics based on microeconomics or, even more, 
based on the ``elementary interactions'' of individuals. 
\par
In order to make some progress in this direction, we will generalize some ideas suggested by
Daganzo to describe the dynamics of supply chains \cite{Daganzo}. 
Like the work by Armbruster {\em et al.} \cite{Armbruster}, 
his approach is related to traffic models as well, but he focusses on models in discrete space in order 
to reflect the discreteness of successive production steps. In order to be able to reflect the economics of
a country, we will have to generalize these ideas to complex production and supply networks. This will be done
in Sec. \ref{Sec1}. Section \ref{Sec2} will, then, focus on business cycles in an sectorally structured economy.
This manuscript can, of course, only be a first step into the direction of
simulating complex production processes or the whole economy of a country. Some further research directions
are indicated in the outlook of Sec. \ref{Sec3}. 

\section{Modelling Supply Networks} \label{Sec1}

{\em The production units:}
We will investigate a system with $u$ production units (e.g. machines or factories) $a, b, c \in \{1,2,\dots,u\}$ 
producing $p$ products $i, j, k\in \{1,2,\dots,p\}$. The respective production process is characterized by 
parameters $n_b^j$ and $f_b^i$: In each production step, production unit $b$ requires
$n_b^j$ products (educts) $j\in \{1,\dots,p\}$ and produces $f_b^i$ products $i \in \{1,\dots,p\}$.
The number of production steps of production unit $b$ per unit time shall be $P_b(t)$. 
It is the product of the processing (departure) rate  $\mu_b$
and the probability $p_b$ of the production unit being occupied:\\[-7mm]
\begin{equation}
 P_b(t) = \mu_b(t) p_b(\rho_b,c_b,s_b) \, .\\[-10mm]
\label{Pb}
\end{equation}
Herein,\\[-8mm]
\begin{equation}
 \rho_b = \lambda_b / \mu_b 
\end{equation}
denotes the channel utilization, $\lambda_b$ the feeding rate,
$c_b$ the number of parallel channels of the production unit, and
$s_b$ the storage capacity. Changing the number $c_b$ of channels or the storage capacity
$s_b$ is costly. For the time being, we will therefore assume $c_b$ to be constant. 
The expression for $p_b(\rho_b,c_b,s_b)$ is well-known
for the stationary state of many queuing systems \cite{queuing}. 
For example, for a $M/M/1:(\infty/\mbox{FIFO})$ process (i.e. one channel with first-in-first-out
serving, unlimited storage capacity, Poisson-distributed arrival times and
exponentially distributed service intervals), one finds 
\begin{equation}
 p_b(\rho_b,1,\infty) = \min \left( \rho_b , 1 \right) \, , % \frac{1 - \rho_b^{s_b - 1}}{1 - \rho_b^{s_b}} , 1 \right) \, .
\end{equation}
so that, in this particular case, we have the relation
%Note that, for $\lambda_b \ll \mu_b$, we have the relation
\begin{equation}
 P_b(t) = \mu_b(t) \min[ \rho_b(t), 1]  = \min[\lambda_b(t),\mu_b(t)] \, . 
\end{equation}
%In other queuing systems, this relation holds approximately under the condition $\lambda_b \ll \mu_b$.
%$P_b$ determines the product flow.
%The feeding rate $\lambda_b$ of the production unit will be specified later.

{\em Feeding rates:}
The feeding rate $\lambda_b$ is not just the product of the concentrations
of educts required for a product, as that would be the case for chemical reactions.
Instead, the feeding rate is determined by the minimum arrival rate $A_{ab}^j$ of required educts $j$,
divided by the number $n_b^j$ of educts required for one production step. 
(The difference is comparable to the Probabilistic AND and the Fuzzy AND in Fuzzy Logic.)
As the arrival rates $A_{ab}^j$ are given by the input buffer flows $\nu_b^j I_b^j$ 
%and of the delivery flows $\sum_a D_{ab}^i$ of product $i$ from other sources $a$ to production unit $b$ 
(where $\nu_b^j$ is the rate of getting educt $j$ from the the input buffer into
the production unit $b$ and $I_b^j$ is the number of educts stored in the input buffer),
the resulting feeding rate is
\begin{equation}
 \lambda_b(t) = \min_j \left( \frac{\nu_{b}^j(t) I_b^j(t)} % + \sum_a D_{ab}^j}
 {n_b^j} \right) \, .
\end{equation}
Note that, due to stochastic variations (which can cause queuing effects),  
efficient production is related to $\lambda_b \le r \mu_b$ with $1 > r \approx 0.7$ \cite{queuing}.
Therefore, it would be reasonable to use transport rates $\nu_b^j$ according to
$\nu_b^j = r \mu_b n_b^j / I_b^j $. However, usually there are capacity constraints $V_{b}^j$ in
the transport of product $j$ in production unit $b$, which may be reflected by a function $U(x,V)$ %with  $U(x,V) \le x$ 
such as 
\begin{equation}
 U(x,V) = \frac{x}{1+x/V} \approx \left\{
\begin{array}{ll}
x & \mbox{ for } x \ll V \\
V & \mbox{ for } x \gg V.
\end{array} \right.
\end{equation}
Therefore, we will make the specification 
\begin{equation}
 \nu_{b}^j(t) = U\!\left(\frac{r \mu_b(t)n_b^j}{I_b^j(t)},V_{b}^j(t)\right) \, . 
\end{equation}
%the equation for the feeding rate becomes
%\begin{equation}
% \frac{d\lambda_b}{dt} = \frac{1}{\tau_\lambda} \left[ \min_i \left( 
%  \frac{\nu_{b}^i I_b^i + \sum_a D_{ab}^i}{n_b^i} \right) - \lambda_b \right] \, , 
%\end{equation}
%as the sum of the input buffer flow $\nu_b^i I_b^i$ and
%of the delivery flow $\sum_a D_{ab}^i$ of product $i$ replaces the delivery from 
%the market $N_b^i$. 
%\par
%If the adaptation to the stationary value takes a considerable adaptation time $T_P$,  
%Eq. (\ref{Pb}) must be replaced by 
%\begin{equation}
%  \frac{dP_b}{dt} = \frac{1}{\tau_P} \left[ \mu_b p_b(\rho_b,c_b,s_b) - P_b \right] \, ,
%\label{Pex}
%\end{equation}

{\em Input and output buffers:}
Let us assume, each production unit $b$ has input buffers for required educts $i$ and output buffers (a warehouse)
for the products. We will assume the input buffers are filled with $I_b^i(t)$ educts
$i \in \{1,\dots,p\}$ and the output buffers with $B_b^i(t)$ products waiting to be delivered.
If $D_{ab}^i(t)$ denotes the delivery flow of products $i$ from production unit $a$ to $b$, the change of
an input buffer stock with time is given by the conservation equation
\begin{equation}
  \frac{dI_b^i}{dt} = \sum_a D_{ab}^i(t) -  n_b^i P_b(t)\, ,
\label{bal1}
\end{equation} 
as $\sum_a D_{ab}^i(t)$ is the number of products $i$ delivered from various sources (production units) $a$ 
and  $n_b^i P_b$ is the number of educts $i$ used up for production per unit time.
Analogously, the dynamics of an output buffer stock is determined by the equation
\begin{equation}
 \frac{dB_b^j}{dt} = f_b^j P_b(t) - \sum_c D_{bc}^j(t) \, ,
\label{bal2}
\end{equation} 
as $f_b^j P_b$ is the number of newly generated products $j$, and $\sum_c D_{bc}^j(t)$ are deliveries
to other production units $c$. The delivery flow $D_{ab}^i$ 
is given by the delivery rate $\nu_{ab}^i$ times the available products $B_a^i$ in the buffer:
\begin{equation}
 D_{ab}^i(t) = \nu_{ab}^i(t) B_a^i(t) \, .
\end{equation}
The delivery flows are adapted to the order flows $O_{ab}^i$. Ideally, one would have
the relation $D_{ab}^i = O_{ab}^i$, but due to capacity constraints $V_{ab}^i$, we will again assume 
\begin{equation}
 \nu_{ab}^i(t) = U\!\left(\frac{O_{ab}^i(t)}{B_a^i(t)},V_{ab}^i(t)\right) \, ,
\end{equation}
which implies $D_{ab}^i \le O_{ab}^i$. 

%However, if a relevant adaptation time $T_\nu$ is required, one should better use
%\begin{equation}
% \frac{d\nu_{ab}^i}{dt} = \frac{1}{\tau_\nu}
% \left[ V\left(\frac{O_{ab}^i}{B_a^i},V_{ab}^i\right) - \nu_{ab}^i \right] \, .
%\end{equation}

{\em Adaptation of capacities:}
One important aspect of production is the adaptation of production capacities. This requires 
usually a lot of time compared to the other time scales of production. The adaptation time $T_b$ 
of the processing rate $\mu_b$ is much greater than the other adaptation times, so that we will
assume the relation
\begin{equation}
 \frac{d\mu_b}{dt} = \frac{1}{T_b} \left[ 
 W\!\left( \sum_j \frac{f_b^j}{B_b^j} ,
 \dots \right) - \mu_b \right] \, .
\label{zwei}
\end{equation}
Herein, $W(y,...)$ is some control function reflecting the strategy by the production management
to adapt the processing rate $\mu_b$, e.g. as a function of the output buffer stocks $B_b^j$ or other variables. 
Normally, it will increase with decreasing output buffer stocks $B_b^j$, but the function $W(y,\dots)$ saturates
due to financial, spatial, or technological limitations and inefficiencies in the processing of high order flows.
In this study, we will assume a function of the form
\begin{equation}
%W(y,\dots) = a \frac{1+b/y}{1+c/y+d/y^2} = a \frac{y^2 + by}{y^2 + cy + d}
W(1/z,\dots) = A \frac{1+Bz}{1+Cz+Dz^2} 
\end{equation}
with $z = 1/y$ and suitable parameters $A$, $B$, $C$, and $D$.
\par
Of course, the transportation capacities $V_b^j$ and $V_{ab}^i$ are adapted as well. 
One may set up separate equations
for this, but for simplicity, we will here assume the proportionality relations
\begin{equation}
  V_b^j(t) = \widehat{V}_b^{j} \mu_b(t) \quad \mbox{and} \quad V_{ab}^i(t) =  \widehat{V}_{ab}^{i} \mu_b(t) \, .
\end{equation}

{\em Order flows, delivery networks, and price dynamics:} 
The flow of orders is basically given by the flow $P_b n_b^i$ of educts
required for the production by unit $b$. 
%but a prefactor $O(I_b^i/n_b^i)$ allows to take into account corrections. 
If the proportions $q_{ab}^i$ with $\sum_a q_{ab}^i = 1$ reflect orders from different producers, we have
\begin{equation}
O_{ab}^i = q_{ab}^i P_b n_b^i \, .
% \frac{dO_{ab}^i}{dt} = \frac{1}{\tau_O} \left[ q_{ab}^i % O\!\left( \frac{I_b^i}{n_b^i}\right)
% P_b n_b^i - O_{ab}^i \right] \, .
\end{equation}
The fractions $ q_{ab}^i$ characterize the delivery or supply network. One can imagine
various scenarios. For example, the fractions could 
be modelled by an evolutionary selection equation with a selection rate $\nu_q$ \cite{evol}: 
\begin{equation}
 \frac{dq_{ab}^i}{dt} = \nu_q \left( F_{ab}^i(t) - \sum_{a'}
  F_{a'b}^i(t) q_{a'b}^i(t) \right) q_{ab}^i(t) \, .
\end{equation}
It makes sense to relate the fitness $F_{ab}^i$ to the inverse
price $p_{ab}^i$ of product $i$, which producer $a$ takes from 
production unit $b$:\\[-8mm]
\begin{equation}
  F_{ab}^i(t) = 1/p_{ab}^i(t) \, .
\end{equation}
There are many pricing strategies, but the common element seems to be
that the price changes according to the law of supply and demand. Therefore, one conceivable
specification would be
\begin{equation}
 p_{ab}^i(t) = p_i^{0}  \, \frac{O_{ab}^i(t)}{D_{ab}^i(t)} \, ,
\end{equation}
where $p_i^0$ is the price when supply $D_{ab}^i$ and demand $O_{ab}^i$ agree. 
\par
Other specifications are, of course, possible as well,
as the above formulas partly depend on the strategies of the human decision makers involved. 
%These strategies are, for example, reflected by the respective function $W(y,\dots)$ for the desired
%processing rate.

\section{Markets and Business Cycles} \label{Sec2}

\begin{figure}[tbph]
\begin{center}
\vspace*{-1cm}
\hspace*{-2mm}\includegraphics[width=4.4cm, angle=-90]{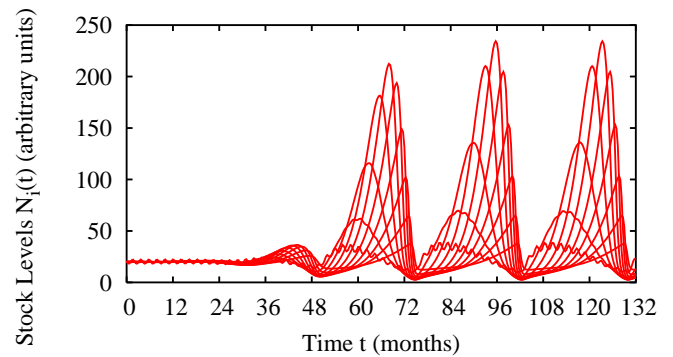}
\end{center}
\caption[]{Variations of the stock levels $N_i(t)$ of $u=p=10$ economic sectors $i$, triggered by a small
perturbation $\delta V_{11}(t) = 0.1 \sin(0.1 t)$ of the constant consumption rate $V_{11}^0 = 0.1 W(1/20)$. 
The model parameters are $A=10$, $B=0.01$, 
$C=0.01$, $D=0.02$, $\widehat{V}_{i}^{i-1}=0.1$, and $T_b = 90$. The initial and boundary conditions are
$N_i(0) = 20 = N_0(t)$, $V_i(0) = \widehat{V}_{i}^{i-1} W(1/N_i(0))$, and $V_{11}(t) = V_{11}^0 + \delta V_{11}(t)$.}
\label{Fig1}
\end{figure}

We will now introduce a simplified and aggregate description of an economy. For this, we will summarize
a whole economic sector by one production unit $b$ and replace input and output buffers by 
markets for the products $i$ of the $u$ different economic sectors. To describe this,
we define the macroscopic stock level of products $i$ in the market by
\begin{equation}
 N_i(t) = \sum_b I_b^i(t) + \sum_b B_b^i(t) \, .
\label{micmac}
\end{equation}
According to the balance equations (\ref{bal1}) and (\ref{bal2}), we find
\begin{equation}
 \frac{dN_i}{dt} = \sum_b (f_b^i - n_b^i) P_b(t) \, .
\label{mac}
\end{equation}
Furthermore, in the arguments of the remaining mathematical relations,
we have to replace input buffer stocks $I_b^i$ and output buffer stocks $B_b^i$ 
by the respective number $N_i$ of products
in the market, as the market replaces the buffers. In this way, we obtain the 
feeding rates
\begin{equation}
 \lambda_b(t) = \min_j \left( \frac{\nu_b^j(t) N_j(t)}{n_b^j} \right) \, ,
\end{equation}
which enter $P_b$ according to equation (\ref{Pb}), the transport rates
\begin{equation}
 \nu_{b}^j(t) = U\!\left(\frac{r \mu_b(t)n_b^j}{N_j(t)},V_{b}^j\right) \, , 
\end{equation}
and the processing rates 
\begin{equation}
 \frac{d\mu_b}{dt} = \frac{1}{T_b} \left[ 
 W\!\left( \sum_j \frac{f_b^j}{N_j(t)} ,
 \dots \right) - \mu_b \right] \, .
\label{eq1}
\end{equation}
For $P_b \approx \lambda_b$, one would have to solve this equation together with
\begin{equation}
 \frac{dN_i}{dt} = \sum_b (f_b^i - n_b^i) \mu_b(t) \min_j \left[ U\!\left(\frac{r n_b^j}{N_j(t)}, \widehat{V}_b^{j}\right)
 \frac{N_j(t)}{n_b^j} \right] \, .
\end{equation}
For a linear supply chain with $n_b^i = \delta_{b,i+1}$ and $f_b^i = \delta_{b,i}$ (where
$\delta_{k,l} = 1$, if $k=l$, otherwise 0), this leads to 
\begin{equation}
  \frac{dN_i}{dt} = V_i(t) N_{i-1}(t) - V_{i+1}(t) N_i(t)  
\label{eq2}
\end{equation}
with
\begin{equation}
  V_i(t) = \mu_i(t) \, U\!\left( \frac{r}{N_{i-1}(t)}, \widehat{V}_{i}^{i-1}\right) \, .
\end{equation}
Interestingly, Eqs. (\ref{eq1}) and (\ref{eq2}) basically
agree with the traffic flow model by Hilliges and Weidlich \cite{Hilliges},
where (\ref{eq2}) is analogous to the equation for the vehicle density and (\ref{eq1}) corresponds to the
velocity equation. These equations are known to behave linearly unstable with respect to perturbations of
$N_i(t)$, if $T_b$ exceeds a certain threshold which depends on the maximum slope 
$dW/dN_i$ \cite{Hilliges}. This is the reason why one can frequently observe an instability of supply chains,
called the ``bullwhip effect'' (e.g. in the ``beer distribution game'' \cite{beergame}).
A thorough analysis shows that the Hilliges-Weidlich model would, in fact, be more suitable for the 
description of supply chains  than of traffic: First, their velocity equation does not contain a convection 
term, as it should be the case for traffic. Second, their equations normally do not show metastability \cite{stopgo}, i.e. 
a region in which traffic flow breaks down, when a perturbation exceeds a critical amplitude,
but where smaller perturbations fade away. For this reason, a supply chain of the above kind is expected to behave 
either stable or linearly unstable, no matter how large the perturbation amplitude is. The most 
interesting point, however, is the reaction of the system to a periodic perturbation in the consumption rate $V_{p+1}(t)$,
as the resulting variations in the stock levels are synchronized and much slower (see Fig.~\ref{Fig1}). For this reason, they
may explain business cycles as self-organized phenomenon.

\section{Summary and Outlook} \label{Sec3}

In this contribution, a theory of supply networks has been sketched, which may open a new area
of econophysics \cite{econophysics}. This theory is developed to help understand the dynamical phenomena,
breakdowns, instabilities and inefficiencies of production processes and supply networks. Future work
will have to address questions such as the relevance of the network structure for the resulting dynamics,
possible control strategies, the role of the market and pricing mechanism, etc. This specifically concerns
the choice of the factor $r$, the specification of the functions $U$, $W$, and the equations suggested
in the paragraph on delivery networks and price dynamics. 
Here, we have focussed on an application to markets, i.e. macroeconomics. For this, we
have replaced input and output buffers by markets for different products [see Eq.~(\ref{micmac})]. 
One may view this as some kind of micro-macro link,
as we have started with equations for single firms (a microeconomic level), e.g. Eqs. (\ref{bal1}), (\ref{bal2}), 
and ended up with equations (\ref{mac}) for
market sectors. Assuming a sectoral structure of economics, one can relate the resulting equations with
the Hilliges-Weidlich model, which has originally been developed for traffic flow. These equations can
describe the ``bullwhip effect'' due to their linear instability in a certain regime of operation. The underlying
mechanism is the slow adaptation of the processing rate to changes in the order flows or stock levels
in the market. Interestingly, the resulting oscillations in the stock levels of the different products $i$ 
have a characteristic frequency, which can be much lower than the underlying fast variations in the consumption
rate. These oscillations synchronize among different economic sectors and may explain business cycles
as a self-organized phenomenon with slow dynamics. As the variables in the model are operational and
measurable, the model can be tested and calibrated with empirical data. Apart from some equations
which were not further applied in this study, most of the proposed model equations were 
conservation equations, equations given by the product flows, or relations derived 
with stochastic concepts used in queuing theory. They reflect the transport and interaction
of products, so that the physics of driven many-particle systems and of complex systems
can make some significant contributions to the new multidisciplinary field of  self-organization phenomena
in production and supply networks.
%Although the production of goods has some analogies with chemical reactions, the
%underlying processes are not identical and require reformulations.
%
%{\em Acknowledgements:} The author would like to thank D. Armbruster, C. Daganzo, A. Kesting, 
%C. K\"uhnert, D. Sanders, T. Seidel, T. Werner, and U. Witt, for inspiring discussions.

\begin{references}
\vspace*{-5mm}
\bibitem{reviews}  T. Nagatani, Rep. Prog. Phys. {\bf 65}, 1331 (2002);
D. Chowdhury, L. Santen, and A. Schadschneider,
%Statistical physics of vehicular traffic and some related systems. 
{Phys. Rep.} \textbf{329}, 199 %--329 
(2000);
D. Helbing, Traffic and related self-driven many-particle
systems. {Rev. Mod. Phys.} \textbf{73}, 1067--1141 (2001).

\bibitem{segregation}
S. B. Santra, S. Schwarzer, and H. Herrmann, 
%\tit{Fluid-induced particle segregation in sheared granular assemblies}
Phys. Rev. E {\bf 54}, 5066 %{5072}
(1996); H. A. Makse, S. Havlin, P. R. King, and H. E. Stanley, 
%\tit{Spontaneous stratification in granular mixtures}
Nature {\bf 386}, 379 %}{382} 
(1997).

\bibitem{colloid}  
%J. Dzubiella and H. L\"owen, 
%%Pattern formation in driven
%%colloidal mixtures: tilted driving forces and re-entrant crystal freezing.
%J. Phys.: Cond. Mat. {\bf 14}, 9383 (2002); 
J. Dzubiella, G. P. Hoffmann, and H. L\"owen, 
%Lane formation in colloidal mixtures driven by an external field. 
Phys. Rev. E {\bf 65}, 021402 (2002).

\bibitem{pedants} D. Helbing and P. Moln\'{a}r, 
%\tit{Social force model of pedestrian dynamics}
Phys. Rev. E {\bf 51}, 4282 %}{4286}
(1995); D. Helbing and T. Vicsek,
%\tit{Optimal self-organization}
New J. Phys. {\bf 1}, 13.1 %}{13.17}
(1999); I. D. Couzin and N. R. Franks,
%Self-organized lane formation and optimized traffic flow in army ants,
Proc. Roy. Soc. London B, 02PB0606.1 %--02PB0606.8
(2002).

\bibitem{phase} D. Helbing, A. Hennecke, and M. Treiber, 
%{Phase diagram of traffic states in the presence of inhomogeneities} 
{Phys. Rev. Lett.} \textbf{82},  4360 %--4363} 
(1999).

\bibitem{panic} D. Helbing, I. Farkas, and T. Vicsek,
%Simulating dynamical features of escape panic,
Nature {\bf 407}, 487 %--490
(2000).

\bibitem{clogging} K. To, P.-Y. Lai, and H. K. Pak,
%Jamming of granular flwo in a two-dimensional hopper.
Phys. Rev. Lett. {\bf 86}, 71 %--74
(2001).

\bibitem{stopgo} B. S. Kerner and P. Konh\"auser, 
%\tit{Structure and parameters of clusters in traffic flow}
{Phys. Rev. E} {\bf 50}, 54 %--83
(1994). %M. Treiber, A. Hennecke, and D. Helbing, 
%%\tit{Derivation, properties, and simulation of a gas-kinetic-based, 
%%non-local traffic model}
%{Phys. Rev. E} {\bf 59}, 239 %--}{253}
%(1999).

\bibitem{Daganzo}
C. Daganzo, {\em A Theory of Supply Chains} (Springer, New York, 2002), in print.

\bibitem{Armbruster}
D. Marthaler, D. Armbruster, and C. Ringhofer,
%%A mesoscopic approach to the simulation of seminconductor supply chains,
in {\em Proc. of the Int. Conf. on Modeling and Analysis of Semiconductor Manufacturing},
ed. by G. Mackulak {\em et al.} (2002), pp. 365; %--369;
%D. Armbruster, D. Marthaler, and C. Ringhofer,
%Modeling a re-entrant factory.
%preprint;
B. Rem and D. Armbruster,
%Control and synchronization in switched arrival systems.
preprint;
I. Diaz-Rivera, D. Armbruster, and T. Taylor,
%Periodic orbits in a class of re-entrant manufacturing systems.
preprint.

\bibitem{business} J. A. Schumpeter,
{\em The Theory of Economic Development}
(Harvard University, Cambridge, MA, 1934);
P. Samuelson,
%Interations between the multiplier analysis and principle of acceleration.
Rev. Econ. Studies {\bf 21}, 75 %--78 
(1939); N. Kaldor,
%A model of the trade cycle,
Econ. J. {\bf 50}, 78 %--92
(1940).

\bibitem{Witt} U. Witt and G.-Z. Sun, 
%Myopic behavior and cycles in aggregate output.
in {\em Jahrb\"ucher f. National\"okonomie u. Statistik}
(Lucius \& Lucius, Stuttgart, 2002), Vol. 222/3, pp. 366. %--376.

\bibitem{queuing} R. Hall, {\em Queueing Methods for Service and Manufacturing}
(Prentice Hall, Upper Saddle River, NJ, 1991); 
T. Saaty, {\em Elements of Queueing Theory with Applications}
(Dover, New York, 1983).

\bibitem{evol} R. Feistel and W. Ebeling,
{\it Evolution of Complex Systems}.
(Kluwer Academic, Dordrecht, 1989).

\bibitem{Hilliges}
M. Hilliges and W. Weidlich, 
%\tit{A phenomenological model for dynamic traffic flow in networks}
{Transpn. Res. B} {\bf 29}, 407 %--431
(1995); M. Hilliges, 
{\it Ein ph{\"a}nomenologisches Modell des dynamischen Verkehrsflusses in Schnellstra{\ss}ennetzen}
(Shaker, Aachen, 1995).

\bibitem{beergame} J. D. Sterman,
{\em Business Dynamics}
(McGraw-Hill, Boston, 2000).

\bibitem{econophysics} R. N. Mantegna and H. E. Stanley, 
{\it Introduction to Econophysics} %: Correlations and Complexity in Finance} 
(Cambridge University, Cambridge, 1999);
J.-P. Bouchaud and M. Potters, 
{\em Theory of Financial Risk} %: From Statistical Physics to Risk Management} 
(Cambridge University, Cambridge, 2000); 
H. Levy, M. Levy, and S. Solomon, 
{\em Microscopic Simulation of Financial Markets}
(Academic, San Diego, 2000).
D. Challet and Y.-C. Zhang, 
%\tit{On the minority game: Analytical and numerical studies} 
{Physica A} {\bf 256}, 514 %--532
(1998). 
\end{references}
\end{document}